# Experimental Confirmation of Massive Dirac Fermions in Weak Charge-Ordering State in $\alpha$-(BEDT-TTF)$_2$I$_3$

Kenta Yoshimura, Mitsuyuki Sato, and Toshihito Osada*

*Institute for Solid State Physics, University of Tokyo, Kashiwa, Chiba 277-8581, Japan.*

The electronic structure of weak charge-ordering (CO) state just below the critical pressure in an organic conductor $\alpha$-(BEDT-TTF)$_2$I$_3$ was experimentally investigated using peak structure in the temperature dependence of interlayer magnetoresistance (MR). Based on a minimal model considering multiple Landau levels (LLs), we discuss herein the MR peak as characteristic to multilayer massless/massive Dirac fermion (DF) systems. MR measured in the weak CO state exhibited a clear MR peak, and its magnetic-field dependence was consistent with the LL behavior of a massive DF with a small gap. Results indicate that the weak CO state in $\alpha$-(BEDT-TTF)$_2$I$_3$ is a massive DF state.

Layered organic conductor $\alpha$-(BEDT-TTF)$_2$I$_3$ has attracted considerable attention because a two-dimensional (2D) massless Dirac fermion (DF) state occurs at high pressures [1]. At ambient pressure, it undergoes a first-order metal–insulator transition into an insulating phase caused by charge ordering (CO) at $T_{CO}$ = 135 K. As schematically illustrated in Fig. 1, the CO phase is suppressed by pressure, and then vanishes at approximately $P_c$ = 1.3 GPa. It was theoretically pointed out that the metallic state above $P_c$ is a massless DF state [2]. After that, it has been experimentally confirmed by magnetoresistance (MR) [3, 4], interlayer Hall effect [5, 6], specific heat [7], and NMR [8] measurements. In the massless DF state, the band dispersion around the Fermi level $E_F$ is regarded as a pair of tilted and anisotropic Dirac cones located at general points in the **k** space.

The interlayer MR in the massless DF phase shows characteristic features under a vertical magnetic field. It shows positive MR at low magnetic fields, which subsequently becomes negative, forming a maximum peak [3]. The negative MR is owing to the increase in degeneracy of the $n$ = 0 Landau level (LL) in the massless DF at the quantum limit, where only $n$ = 0 LL is partially occupied [4]. Conversely, the positive MR at low magnetic fields is because of the LL mixing owing to the non-vertical (tilting) interlayer transfer [9]. In this case, the interlayer tunneling probability depends on the vertical field, resulting in positive MR. The MR peak in the field dependence corresponds to the crossover from positive MR to negative MR.

In this study, we focused on "weak CO" state just below $P_c$, where the CO transition temperature $T_{CO}$ is largely suppressed, and the specific latent heat seems small like a second-order transition. In the weak CO state, we can expect a quasi-particle with 2D massive DF nature, that is, the dispersion where a small CO gap opens up at the Dirac points of two Dirac cones (valleys), as shown in the inset of Fig. 1. The angle-dependent interlayer MR under in-plane fields, which shows continuous change between the weak CO and massless DF states [10], supports this expectation. Because the CO gap breaks the inversion symmetry, there are two types of CO domains. In the weak CO state, excess metallic conduction is often observed despite a finite gap [10, 11]. This may originate from conduction along domain boundaries, which has been theoretically discussed based on the massive DF picture [12, 13]. In the massive DF state, various topological properties are expected owing to their finite Berry curvature. The valley Hall effect in the trivial insulator phase [12] and the anomalous Hall effect in the Chern insulator phase [14] were theoretically discussed. Furthermore, the nonlinear anomalous Hall effect has been investigated in the current-carrying state in the weak CO state [15].

However, the massive DF in the weak CO state has not been experimentally confirmed thus far. Therefore, for confirmation, we employed the indirect method, which Sugawara *et al.* originally used to confirm the massless DF in $\alpha$-(BEDT-TTF)$_2$I$_3$ above $P_c$ [16]. They noticed the peak structure in the temperature dependence of the interlayer MR, and demonstrated that the field dependence of the peak temperature, $T_{max}$, reflects the LL of the massless DF. In this paper, we first provide the theoretical basis of the the method by Sugawara *et al.* [16], and extend it to a massive case. We provide a minimal model for the MR peak in temperature dependence, and show that the peak structure is characteristic to multilayer DF systems. Thereafter, we show the experimental results on the interlayer MR in the weak CO state in $\alpha$-(BEDT-TTF)$_2$I$_3$, and experimentally confirm the massive DF using the method by Sugawara *et al*.

First, we present a minimal model which can reproduce the peak structure in the temperature dependence of interlayer MR in multilayer DF systems. In this model, we consider multiple LLs, but ignore the temperature dependence of scattering (mobility), the scattering broadening of LLs, and the tilting effect of the non-vertical interlayer transfer which causes interlayer tunneling between different LLs. We employ the following model of the multilayer system, where 2D massive DF layers stack with weak vertical interlayer transfer.

$$H^{(\pm)}(\mathbf{k}) = \pm\gamma k_x \sigma_x + \gamma k_y \sigma_y + \Delta \sigma_z - 2t_c \cos ck_z \sigma_0, \tag{1}$$

where $(\sigma_x, \sigma_y, \sigma_z)$ are the Pauli matrices, and $\sigma_0$ is a $2\times 2$ unit matrix. The double sign $(\pm)$ indicates two Dirac cones, $(+)$ valley and $(-)$ valley, which form a Kramers pair. $\gamma$, $\Delta$, $t_c$, and $c$ are the parameters that represent the in-plane velocity, mass gap generated by CO, vertical (non-tilting) interlayer transfer, and interlayer distance, respectively. Here, we employ the tunneling picture, which treats the interlayer coupling, $H_\perp \equiv -2t_c \cos ck_z \sigma_0$, as a perturbation [17]. This treatment is correct when the interlayer coupling is incoherent, that is, $t_c \ll \hbar/\tau$, where $\tau$ is the in-plane scattering time. The non-perturbed states are the $n$-th LL states of the $(\pm)$ valley on each 2D layer, and their energies are expressed as follows: $E_n^{(\pm)} = \text{sgn}(n)\sqrt{(2\gamma^2/l^2)|n| + \Delta^2}$ ($n \neq 0$) and $E_{n=0}^{(\pm)} = \mp\Delta$, where $n$ is an integer and $l^2 \equiv \hbar/e|B_z|$. We ignore the Zeeman splitting of LLs for simplicity, resulting in two-fold spin degeneracy in each LL. The 0-th LL shows valley splitting, whereas the other LLs have a two-fold valley degeneracy. The magnetic-field dependence of LLs is depicted in Fig. 2(a). Under a vertical magnetic field, $\mathbf{B} = (0, 0, B_z)$, the matrix element of the interlayer coupling, $H_\perp$, between the LL states on different layers becomes non-zero only when the 2D quantum numbers of the two states are equal and their layers are neighboring [6, 9]. Simply, no LL mixing occurs during interlayer tunneling in the case of vertical transfer [9]. The lowest order contribution of $t_c$ to the interlayer conductivity $\sigma_{zz}$ under the vertical magnetic field is obtained by using the tunneling picture for interlayer transport as follows [17].

$$\sigma_{zz} = 2\frac{t_c^2 c e^3 \tau}{\pi \hbar^3} \frac{|B_z|}{k_B T} \sum_{\pm} \sum_{n=-\infty}^{\infty} f(E_n^{(\pm)})\{1 - f(E_n^{(\pm)})\}, \qquad (2)$$

where $\tau$ is the constant in-plane scattering time and $f(E) = 1/[1 + \exp\{(E-\mu)/k_B T\}]$ is the Fermi distribution function. Here, we assumed that the thermal distribution width $k_B T$ is significantly larger than the LL broadening owing to in-plane scattering ($k_B T \gg \hbar/\tau > t_c$). The interlayer MR is proportional to $\rho_{zz} = 1/\sigma_{zz}$.

This formula achieves a remarkable temperature dependence characteristic of multilayer DF systems. As illustrated in Fig. 2(b), the interlayer MR exhibits a peak structure indicated by an arrow whose position depends on the magnetic field. At low temperatures, the MR exhibits insulating behavior owing to thermal activation onto the CO gap. The insulating behavior disappears in the massless DF case ($\Delta = 0$).

The temperature dependence of the interlayer MR is caused by the thermal broadening of $(1/k_B T)f(E)\{1-f(E)\}$ in Eq. (2). The chemical potential $\mu$ is constantly located at zero energy because of the electron-hole symmetry. The energy region with finite $f(E)\{1-f(E)\}$, which we refer to as the active region below, linearly expands with temperature, and it is represented by $|E - \mu| < k_B T / \alpha$ with a scale factor $\alpha$. (i) At low temperatures ($k_B T \ll |\Delta|$), where the active region is confined between $E_0^{(+)} = -\Delta$ and $E_0^{(-)} = \Delta$, the interlayer MR shows an insulating behavior. (ii) As the temperature increases, $E_0^{(+)}$ and $E_0^{(-)}$ levels enter the active region, saturating $\sigma_{zz}$;

thus, resulting in the MR minimum. (iii) In the temperature region where only $E_0^{(+)}$ and $E_0^{(-)}$ levels are located in the active region, the interlayer MR increases linearly owing to the factor $1/k_B T$ in Eq. (2). (iv) When the active region reaches $E_1^{(\pm)}$ and $E_{-1}^{(\pm)}$ as shown in Fig. 2(a), the interlayer MR decreases owing to the contribution of the $n = \pm 1$ LLs, showing an MR peak. (v) Above the peak temperature, other LLs sequentially begin to contribute to transport, and the MR monotonously decreases, because the number of LLs in the active region increases faster than $k_B T$ owing to non-equidistant LLs in the 2D DF system with linear energy dependence of density of states (DOS).

The peak structure in the temperature dependence of the interlayer MR is a characteristic phenomenon in multilayer DF systems. This originates from the large energy separation between the 0-th LL ($E_0^{(\pm)}$) and the 1-st LL ($E_{\pm 1}^{(\pm)}$) as well as the non-equidistant LLs. In contrast, the peak structure cannot be expected in non-Dirac multilayer systems with constant LL spacing. In fact, in a multilayer system consisting of non-Dirac semiconducting layers, we can demonstrate that the interlayer MR is almost temperature independent with no peak except for the insulating behavior. Therefore, the MR peak can be used for experimental confirmation of 2D massless/massive DFs in multilayer systems.

The peak temperatures ($T_{max}$) obtained from Fig. 2(b) are also plotted in Fig. 2(a). They are well fitted by $k_B T_{max} = \alpha(E_1^{(\pm)} - \mu)$ as indicated by a dashed curve, and the fitted value of the scale factor $\alpha = 0.28$. This confirms that the MR peak appears when the upper edge ($k_B T / \alpha$) of the active region reaches $E_1^{(\pm)}$. Therefore, we can obtain the DF parameters, $\gamma$ and $\Delta$, by fitting $k_B T_{max} = \alpha(E_1^{(\pm)} - \mu)$ to the experimental data $\left(\left|B_z\right|, T_{\max}\right)$. Sugawara *et al.* performed this fitting using $E_1^{(\pm)}$ with a fixed $\Delta = 0$ for case of massless DFs [16]. Therefore, the theoretical basis of this method has been provided, and also its general applicability to massless/massive DF systems has been confirmed.

Here, we note that the MR peak in the temperature dependence is a more general phenomenon than the peak between positive and negative MR regions in the field dependence, because the former can be reproduced without any non-vertical transfer, but the latter cannot. The present model, which considers only a vertical interlayer transfer, cannot derive the low-field positive MR and MR peak in the field dependence. Therefore, the MR peak in the temperature dependence is not necessarily the same as the MR peak in the field dependence; thus, contradicting Sugawara's original idea [16].

Using Sugawara's method, we experimentally investigated whether the weak CO state in $\alpha$-(BEDT-TTF)$_2$I$_3$ is a massive DF state. Single crystals of $\alpha$-(BEDT-TTF)$_2$I$_3$ were synthesized by standard electrochemical oxidation. The typical size of the crystal was $0.15 \times 0.1 \times 0.02$ mm$^3$, and the current and voltage contacts are fabricated using carbon paste on the top and bottom surfaces of crystal. Hydrostatic pressures up to 1.6 GPa were applied using a clamp-type pressure cell, and the pressure was monitored by the resistance of the manganin wire at room temperature. The interlayer MR was measured by the dc four-terminal method using a 10 T superconducting magnet with a $^4$He cryostat (2–300 K) and a rotation mechanism. The CO transition temperatures of the measured samples are plotted in Fig. 1.

Figure 3(a) shows the interlayer MR in the weak CO state ($P = 1.12$ GPa) and massless DF state ($P = 1.58$ GPa) of the same $\alpha$-(BEDT-TTF)$_2$I$_3$ crystal. As for the massless DF state, the previously reported features [3, 4], that is, positive MR in low fields $B < 0.05$ T, negative MR in $0.05$ T $< B < 1$ T, and positive MR in $B > 1$ T originating from the spin polarizing of the $n=0$ LL were well reproduced. Although the MR in the weak CO state is approximately 10 times larger than that in the massless DF state at low magnetic fields, the MR in the weak CO state exhibits qualitatively similar behaviors as the massless DF state. Furthermore, we measured the dependence of MR on the magnetic-field orientation, which is measured by the elevation angle $\theta$, in the weak CO state (1.12 GPa), as shown in Fig. 3(b). Solid curves are the fitting curves of $R_{zz} \propto \left(\left|B_z\right|+B_0\right)^{-1}$ with $B_0$ = 4.6 T, following the case of the massless DF state [3]. Moreover, the angle dependence shows qualitatively similar behavior to that in the massless DF state. The absence of the coherence peak in the parallel field directions ($\theta = 0°$, $180°$) justifies the incoherent interlayer coupling of the system. These results suggest that the DFs also exist in the weak CO state.

To confirm the massive DFs in the weak CO state experimentally, we measured the temperature dependence of the interlayer resistance $R_{zz}(B)$. The inset of Fig. 4(a) shows the temperature dependence of the interlayer resistance $R_{zz}(B)$ at several vertical magnetic fields in the weak CO state ($P$ = 1.12 GPa). It is remarkable that the temperature dependence of MR at each magnetic field shows a clear peak structure, whose position ($T_{max}$) depends on the magnetic field. In addition, the MR monotonously decreases at temperatures higher than $T_{max}$. These behaviors are expected in the multilayer DF system as shown in Fig. 2(b), but never in non-Dirac systems, as aforementioned.

Following Sugawara's work, we show the normalized MR $\{R_{zz}(B)-R_{zz}(0)\}/R_{zz}(0)$ in Fig. 4(a), where the effect of temperature dependence of scattering is removed. The observed result at each magnetic field is qualitatively reproduced in Fig. 2(b). However, comparison between the data from different fields disagrees with Fig. 2(b). In the temperature range above $T_{max}$ in Fig. 4(a), the interlayer MR increases with magnetic fields; simply, it shows positive MR, which is in contrast with the negative MR in Fig. 2(b). This disagreement is because the present model cannot achieve positive MR without any non-vertical transfer. Since the positive MR is caused by the $n \neq 0$ LLs and appears only in the high-temperature side of the peak, the peak temperature $T_{max}$ is expected not to be affected by the positive MR.

The magnetic-field dependence of the MR peak temperature $T_{max}$ is plotted in Fig. 4(b). The inset shows the field dependence of LLs of the massive DF with spin splitting owing to the Zeeman effect. We attempted the fitting of the following formula considering spin splitting.

$$k_B T_{max} = \alpha \left| E_{1\uparrow}^{(+)} - \mu \right| = \alpha \left| \sqrt{(2\gamma^2/l^2)|n| + \Delta^2} - \frac{g}{2}\mu_B |B_z| - \mu \right|. \quad (3)$$

Here, $E_{1\uparrow}^{(+)} = E_1^{(+)} - (g/2)\mu_B |B_z|$ is the lower spin-split level of the $n = 1$ LL, where $g = 2$ is the g-factor, and $\mu_B$ is the Bohr magneton. The fitting curve of Eq. (3) is indicated by the solid curve labeled "massive DF" in Fig. 4(b), and it is in good agreement with the experimental data. For comparison, the fitting result assuming the massless DF (with a fixed $\Delta = 0$) is also shown by the dashed curve in Fig. 4(b), which obviously deviates from the data.

We obtained $\alpha = 2.3$, $\gamma/\hbar = 0.92 \times 10^4$ m/s, and $|\Delta| = 0.17$ meV as the values of the fitting parameters. The fitted value of $\alpha$ is rather larger than that expected from the present model, and the value of $|\Delta|$ is much smaller than the CO transition temperature $T_{CO}$~25K defined by the peak of $-d\log R_{zz}/dT$ (Fig. 1). These disagreements are considered to originate from the scattering broadening of LLs. If $E_{1\uparrow}^{(+)}$ in Eq. (3) can be substituted by $E_{1\uparrow}^{(+)} - \Gamma$ in the presence of finite LL width $\Gamma$, the fitted $E_{1\uparrow}^{(+)}$ and $\alpha$ will be underestimated and overestimated, respectively. In this sense, the present fitting results include large quantitative ambiguity. However, the field dependence of $T_{\max}$ is expected to reflect that of $E_{1\uparrow}^{(+)}$ since the scale factor $\alpha$ is constant. Therefore, we use the results only for the qualitative confirmation of massive DFs. Because $\Gamma$ depends on the temperature, magnetic field, and the line-shape of broadening in real systems, additional assumptions beyond the present model are necessary to discuss the effect of level broadening.

The present results strongly suggest that the quasi-particles in the weak CO state have a massive DF dispersion, in which a small gap opens at the Dirac point of the massless DF state. Because the space inversion symmetry is broken in the weak CO state, there exist two types of CO domains with opposite signs of $\Delta$, which are connected by the inversion operation. In the weak CO state, the transport properties could be affected by the metallic transport along the domain boundary [9, 10], resulting in strong sample dependence. Nevertheless, the MR peak temperature is considered to be unaffected by the domain effect, because both types of domains have the same LL configuration.

In conclusion, we have investigated the electronic structure of the weak CO state just below $P_c$ in an organic conductor $\alpha$-(BEDT-TTF)$_2$I$_3$, using the peak structure in the temperature dependence of interlayer MR. The MR peak can be derived considering multiple LLs and only vertical interlayer transfer. It is a phenomenon characteristic of multilayer massless/massive DF systems, and its magnetic-field dependence provides information on the DF systems. The interlayer MR measured in the weak CO state exhibited a clear MR peak in the temperature dependence, and the field dependence of the peak was consistent with the LL behavior of the massive DF with a finite gap. This provides indirect experimental evidence that the weak CO state in $\alpha$-(BEDT-TTF)$_2$I$_3$ is a massive DF state with a finite gap. This means that various topological effects caused by finite Berry curvature can be expected in the weak CO state in $\alpha$-(BEDT-TTF)$_2$I$_3$. In addition, some critical behavior might appear around the critical pressure [18].

The authors thank Dr. G. Matsuno and Prof. N. Tajima for useful discussions. They also thank Ms. A. Mori and Dr. T. Konoike for growing single crystals, Dr. K. Uchida for technical support, and Dr. T. Taen for valuable discussions. This work was supported by JSPS KAKENHI Grant Numbers JP25107003 and JP20H01860.

**Figure 1** (Yoshimura *et al.*)

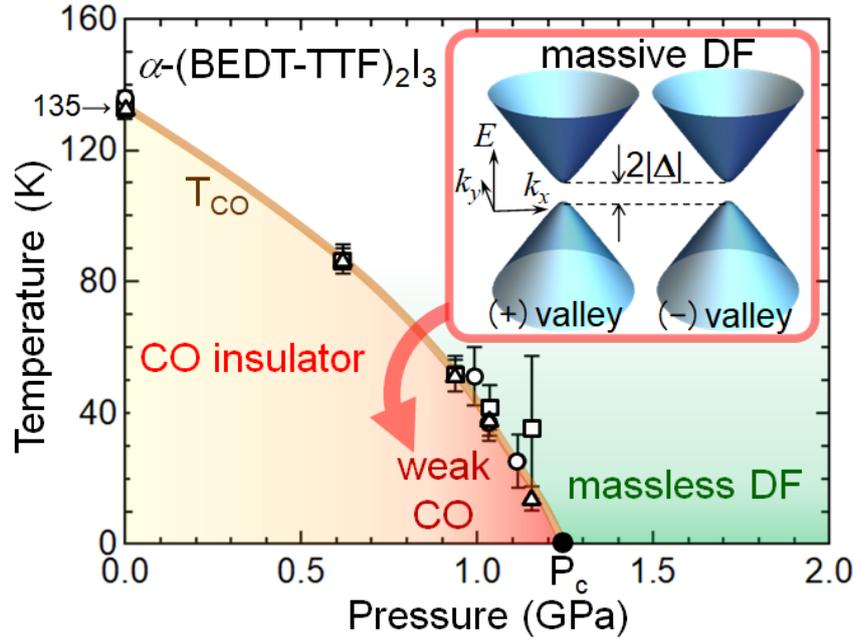

**Fig. 1.** (color online)

Schematic phase diagram of $\alpha$-(BEDT-TTF)$_2$I$_3$. The CO insulator region just below the critical pressure, $P_c$, is referred to as the weak CO state. The circle, square, triangle symbols indicate the transition temperatures of the three samples measured in the present work. They were determined as the peak temperature of $-d\log R_{zz}/dT$ following Ref. 11. The inset illustrates the 2D massive DF dispersion considered in the present model.

**Figure 2** (Yoshimura *et al.*)

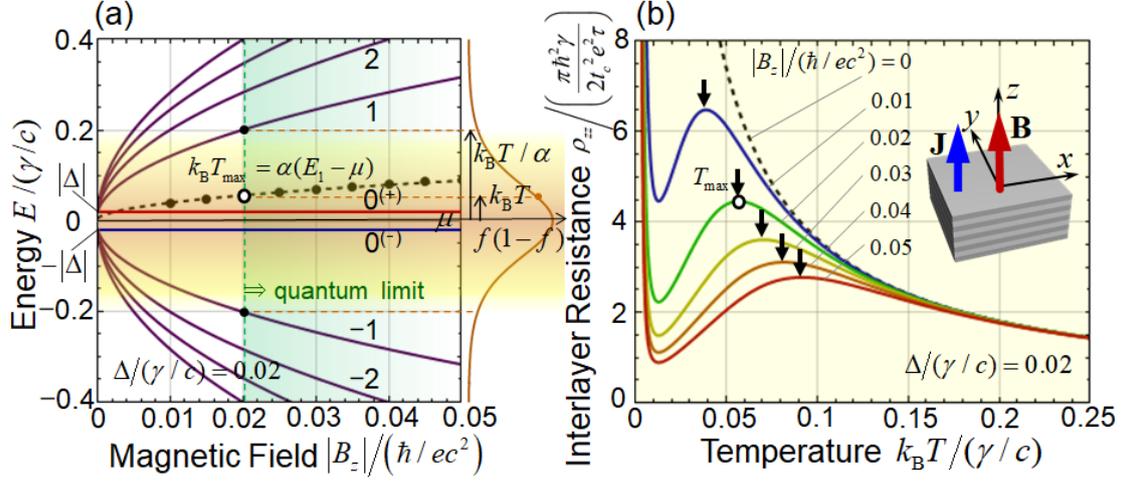

**Fig. 2.** (color online)

(a) Magnetic-field dependence of LLs of massive DFs with no spin splitting. The *n* = 0 LL shows the valley splitting into two levels, labeled "0$^{(+)}$" and "0$^{(-)}$". The active region of thermal distribution $f(E)\{1- f(E)\}$ is represented by the background gradation. The MR peak temperatures are also plotted. (b) Interlayer MR in the multilayer massive DF system calculated as a function of temperature under several vertical magnetic fields. The dashed curve indicates the temperature dependence of the interlayer resistance at zero magnetic field. The inset shows the configuration of the model.

**Figure 3** (Yoshimura *et al.*)

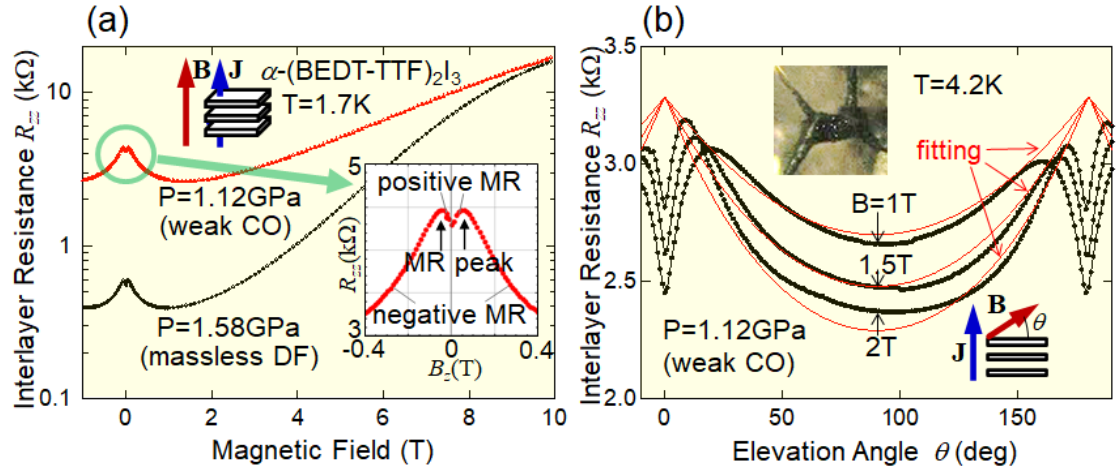

**Fig. 3.** (color online)

(a) Magnetic-field dependence of the interlayer resistance in the weak CO state (1.12 GPa) and massless DF state (1.58 GPa) of $\alpha$-(BEDT-TTF)$_2$I$_3$ at 1.7 K. The inset shows the weak-field region in enlarged scale. (b) Dependence of interlayer resistance on the elevation angle of the magnetic field in the weak CO state (1.12 GPa) at 4.2 K. The microscope image of the sample is presented. In both figures, the experimental configuration is also illustrated as insets.

**Figure 4** (Yoshimura *et al.*)

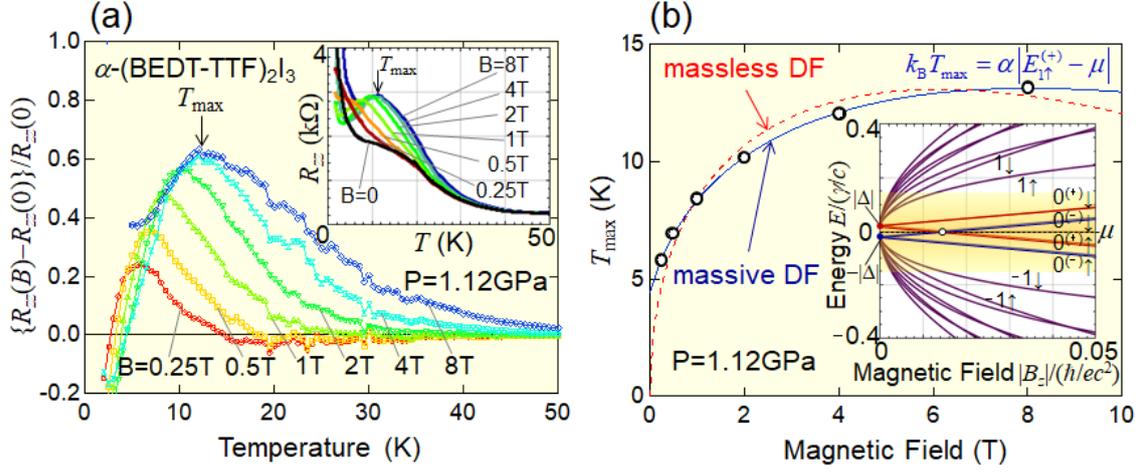

**Fig. 4.** (color online)

(a) Temperature dependence of the normalized interlayer MR in the weak CO state (1.12 GPa) at several magnetic fields normal to the conductive plane. Inset: temperature dependence of the raw interlayer resistance. (b) Magnetic-field dependence of the MR peak temperature $T_{\max}$. The solid curve is the fitting result using Eq. (3), and the dashed curve is the fitting result fixing $\Delta = 0$. The inset illustrates the LLs of the massive DF dispersion with spin splitting.

17